# Beyond Subjectivity: Continuous Cybersickness Detection Using EEG-based Multitaper Spectrum Estimation

Berken Utku Demirel ⓘ, Adnan Harun Dogan ⓘ, Juliete Rossie ⓘ, Max Möbus ⓘ, and Christian Holz ⓘ

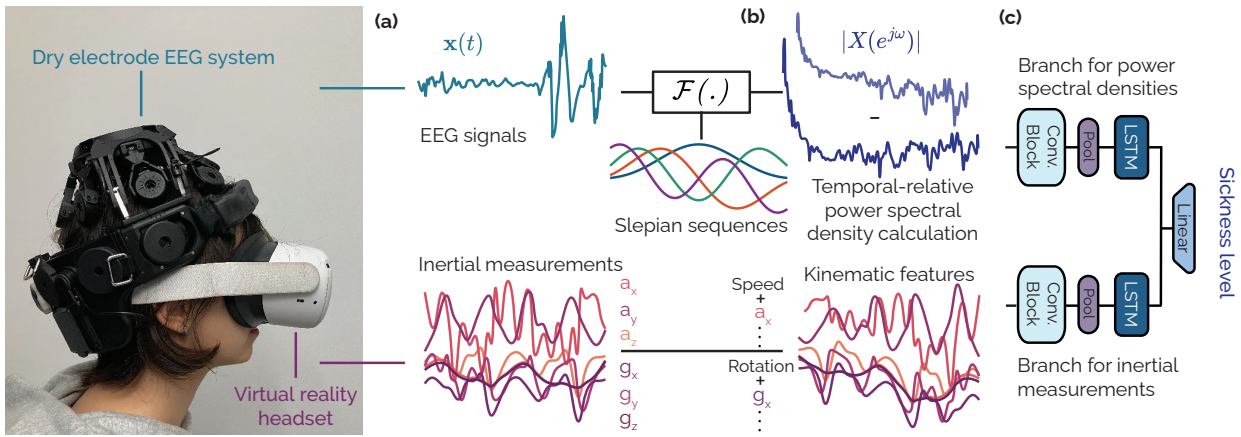

Fig. 1: Our proposed method continuously estimates levels of cybersickness from (a) EEG signals and head motions in VR. (b) Our multitaper-based approach extracts suitable representations from the EEG and inertial signals. (c) The two encoders in our neural network take representations of EEG and inertial signals as input, respectively, to independently extract modality-specific features from the spectral density before concatenating them to estimate a user's sickness level.

**Abstract**— Virtual reality (VR) presents immersive opportunities across many applications, yet the inherent risk of developing cybersickness during interaction can severely reduce enjoyment and platform adoption. Cybersickness is marked by symptoms such as dizziness and nausea, which previous work primarily assessed via subjective post-immersion questionnaires and motion-restricted controlled setups. In this paper, we investigate the *dynamic nature* of cybersickness while users experience and freely interact in VR. We propose a novel method to *continuously* identify and quantitatively gauge cybersickness levels from users' *passively monitored* electroencephalography (EEG) and head motion signals. Our method estimates multitaper spectrums from EEG, integrating specialized EEG processing techniques to counter motion artifacts, and, thus, tracks cybersickness levels in real-time. Unlike previous approaches, our method requires no user-specific calibration or personalization for detecting cybersickness. Our work addresses the considerable challenge of reproducibility and subjectivity in cybersickness research. In addition to our method's implementation, we release our dataset of 16 participants and approximately 2 hours of total recordings to spur future work in this domain.
Source code: `https://github.com/eth-siplab/EEG_Cybersickness_Estimation_VR-Beyond_Subjectivity`.

✦

## 1 INTRODUCTION

Virtual reality (VR) provides immersive experiences that enable users to engage in diverse activities, from entertainment to healthcare [53, 57]. VR can simulate environments and scenarios with high fidelity, allowing users to engage in activities that mimic real-world interactions [6, 14]. From training simulations for medical professionals to therapeutic interventions for patients with phobias or post-traumatic stress disorder (PTSD), VR can effectively support experiential learning and treatment [20, 38, 58].

While VR offers promising benefits across various fields, a significant obstacle is the prevalence of cybersickness among individuals. The lack of means for cybersickness mitigation poses a barrier to the widespread adoption and long-term usability of VR devices. Cyber-

• *The authors are with the Department of Computer Science at ETH Zürich, Zurich, Switzerland.*
• *E-mail addresses: firstname.lastname@inf.ethz.ch.*

sickness can manifest as dizziness, nausea, stomach discomfort, and burping, and can last up to a week [55]. As VR technologies continue to advance, there is a need for effective methods to detect, monitor, and not least prevent cybersickness from occurring to begin with. Thus, continuously monitoring participants' sickness levels is crucial for enhancing user experience and satisfaction but also for ensuring safety and well-being in VR. By responding to changes in cybersickness levels in real-time, interventions could promptly mitigate adverse effects.

Previous cybersickness studies often collected subjective measures of cybersickness only after the VR immersion [5, 60]. Such approaches cannot estimate the continuous and immediate changes in perceived sickness levels, which would be crucial for facilitating interventions to prevent symptoms of cybersickness from occurring [1]. Many existing approaches, often stereo-image-based, further assume that the same sequence of frames and sounds always induces the same effect on different participants [35]. However, cybersickness remains highly subjective, and individuals exhibit varying reactions in response to the same frame sequence in VR [25, 47].

More sophisticated approaches have collected bio-physiological measurements to estimate cybersickness [11, 27, 42] while restricting participants' movement within VR to minimize motion artifacts and acquire clean biosignals [69]. They showed that the features extracted from these measurements (e.g., electrodermal activity (EDA), heart rate (HR), pupil size, and brain activity) correlate with participants' reported cybersickness levels. Restricting motions is especially relevant for recording clean EEG signals, which have shown promising for cybersickness detection [35, 43]. Unfortunately, such movement and interaction constraints are not representative of user behavior in VR [66]. Therefore, there is a need for more robust approaches for continuous cybersickness detection that work in the presence of motion.

In this paper, we first introduce a novel dataset of EEG recordings from 16 participants during cycling simulations in VR. We then propose a method for continuously estimating cybersickness levels in real time, with minimal latency, enabling timely interventions without the need for user-specific fine-tuning.

Our work underscores the importance of continuous monitoring of cybersickness for optimizing the VR experience. To increase opportunities of reproducibility and generalizability for future work, we release the implementation of the method as well as our 16-participant dataset for public access.

**Contributions**

In summary, we make the following contributions in this paper:

- a novel processing approach to extract meaningful features from EEG signals to continuously estimate cybersickness in VR. Unlike traditional methods, our modified multitaper-based method leverages a superior spectral resolution and reduced variance while considering the temporality in EEG, which allows it to accurately and reliably predict cybersickness levels. We validated the energy and memory efficiency of our approach on real hardware.

- an interpretation of the EEG power spectral density characteristics with physiological markers commonly used in neurophysiology for different levels of sickness. This demonstrates the insightful link between the effect of cybersickness on brain activity and established methods in neurophysiological research.

- an extensive evaluation of our method, including ablation studies on a dataset of participants interacting in VR, accompanied by EEG and head-motion recordings. Our method reaches or outperforms the current state-of-the-art methods for detecting and estimating cybersickness on the same dataset while reducing the required number of input modalities.

- a public release of our dataset, method implementation, and evaluation analysis. Our ablation study includes multiple models to support future research and experimentation. Our dataset will facilitate replication and validation of our findings by the research community.

## 2 RELATED WORK

There is a large volume of published studies seeking the reasons and detection of cybersickness in VR. Theories regarding cybersickness include sensory conflict theory, poison theory, and postural instability theory [9, 39]. While the existing literature is extensive and focuses particularly on the sensory conflict theory which claims that pseudo-motion-perception, perceived by human visual stimuli, is responsible for cybersickness when the individual is stationary in reality, we review a wide range of cybersickness research.

### 2.1 Stereo-image based sickness detection

Prior researchers have investigated the effectiveness of using stereo-image datasets collected from videos to predict cybersickness [33, 49, 52]. For example, Padmanaban et al. [52] used 19 two-minute VR videos with depth and optical flow features to predict cybersickness. Later, 3D convolutional neural networks are proposed to detect sickness with a multimodal deep fusion approach for optical flow, disparity, and saliency features [30]. Similarly, Kim et al. [28], used a convolutional auto-encoder [48] to predict cybersickness by utilizing reconstruction error captured from exceptional motion videos. However, the videos used in most of the prior research were pre-recorded and were rendered using the HMDs, instead of allowing the participants to interact in the VR simulations. For example, Kim et al. [32] used the KITTI dataset [17], which are not VR videos. Padmanaban and Lee et al. used pre-recorded short VR videos (1–2 minutes long), which also did not allow free locomotion and may not have included a long enough exposure to introduce cybersickness [50, 54]. Jin et al. [29] used five different VR videos and allowed users to do different types of locomotion and 3D-object manipulation and achieved a coefficient of determination (R2) value of 86.8% in predicting cybersickness from the video features.

While the aforementioned methods achieved comparable results by leveraging the sequences of frames from stereo-image datasets for predicting cybersickness, it is important to note that the designed models are often computationally expensive and demanding. Especially, the intricate architectures and processing requirements of neural networks used in these approaches demand substantial computational resources, hindering their practical utility for immediate cybersickness detection during VR experiences. This computational complexity poses challenges in deploying these models for the real-time detection of cybersickness. Considering this crucial limitation of previous works, our proposed method is lightweight, as it does not rely on video data, enabling its deployment in real-time scenarios for the prediction of cybersickness. This capability facilitates timely interventions to mitigate discomfort and adverse effects experienced by users.

### 2.2 Physiological signal-based sickness detection

Although stereo-image-based sickness detection methods yield competitive results, they lack generalizability due to the fact that individuals can exhibit different reactions to identical sequences of frames within virtual environments [8]. In other words, a sequence of frames inducing sickness in one person may stimulate no response in another. Hence, recent studies have turned to bio-physiological measures for predicting cybersickness [11, 26, 27] and showed that there is a significant correlation between cybersickness and changes in the bio-physiological signals. For example, researchers have identified a significant positive correlation between cybersickness and HR and EEG delta waves, as well as a negative correlation with EEG beta waves [36, 43]. Additionally, some studies reported that galvanic skin response (GSR) on the forehead has a higher correlation with cybersickness and could be used to predict cybersickness [16, 56, 70]. Similarly, heart rate and GSR information are combined to predict cybersickness with an accuracy of 87.38% using neural networks that were collected from 22 participants [23]. Also, Kim et al. collected 8-channel EEG data from 200 participants immersed in 44 different VR simulations to detect cybersickness [35]. Most of these cybersickness studies involving physiological signals are confined to seated conditions with restricted locomotion, aiming to mitigate motion artifacts and noise during data collection [3, 24, 69]. This limitation is particularly significant for EEG signals—one of the primary physiological signals for cybersickness detection—given its exposure to sensory conflicts.

However, in this paper, we present a novel preprocessing step for cybersickness detection designed to effectively eliminate noise and motion artifacts from EEG signals. This preprocessing step allows for the deployment of our proposed method in realistic VR environments without restricting the motions of users. By addressing the challenge of motion artifacts, our method enhances the reliability and accuracy of cybersickness detection using physiological signals. Furthermore, our approach contributes to bridging the gap between laboratory-based studies conducted under controlled conditions and real-world applications where users have unrestricted movement within virtual environments. This advancement marks a significant step towards the development of practical and effective solutions for cybersickness detection using EEG signals and mitigation in immersive experiences.

### 2.3 Motion-based sickness detection

Due to the limitations of the cybersickness prediction from stereoscopic video and physiological signals, recent research has been focused on predicting cybersickness from headsets using the eye-tracking, and motion data [7, 13, 19, 22, 44]. For example, Chang et al. reported that different eye features (e.g., fixation duration and distance between the eye gaze and the object position sequence) are highly correlated when the participants felt cybersickness and proposed and support vector

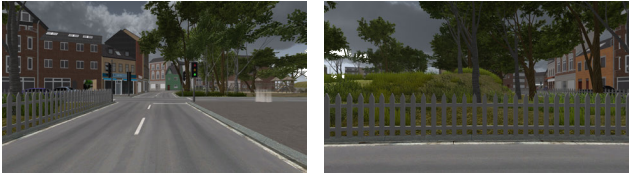

(a) An example of VR scene for users while going front

(b) The side view of the scene to have continuity during VR immersion

Fig. 2: The example views from the designed virtual environments with front (Figure 2a) and side (Figure 2b).

machine (SVM) regression for cybersickness prediction [7]. Similarly, Lopes et al. [44] reported that pupil position and eye-blink rate between the sickness group and the non-sickness groups were significantly different. There is also a large volume of published studies describing the role of using head-tracking and postural data for cybersickness prediction [2, 13, 54, 67]. In addition to using objective measurements (i.e., biophysiological signals, stereoscopic video, inertial measurements) for cybersickness studies, researchers often use subjective measures to detect cybersickness severity. The most commonly used cybersickness subjective measures are simulator sickness questionnaire (SSQ) [55]. However, several researchers have argued that cybersickness is different from simulator sickness [5, 60] and proposed cybersickness susceptibility questionnaire [15] and virtual reality sickness questionnaire [34] for subjective measurement of cybersickness. Yet, these subjective measures are often collected after the VR immersion. Therefore they do not provide sufficient granular understanding of cybersickness severity during VR immersion.

Thus, in this paper, we employed a setup that allowed for continuous reporting of sickness levels through a joystick controller, similar to [42]. We also present a novel approach for cybersickness prediction from the EEG signals using a tailored preprocessing technique for the VR environments, which allowed us to use EEG signals under noisy conditions. We had significant detection accuracy by using the proposed approach, which we believe can be used to develop a standalone cybersickness prediction framework. More importantly, our proposed method can continuously estimate and follow the degree of cybersickness level of users in the virtual environment instead of waiting for minutes to get a response from the users.

## 3 DATA COLLECTION

### 3.1 Virtual Environments

We designed the virtual environment (see Figure 2 for an example) for cybersickness detection using Unity 3D to maximize a realistic VR immersion for accurate data collection. The user interfaces (UIs) were strategically positioned based on participants' height and arm length to optimize usability. To enhance participants' sense of agency, we simplified their hand and controller representations to be transparent white spheres. Before each session, participants confirmed the alignment of virtual representations with their physical movements, ensuring calibration accuracy. Importantly, the virtual environment's elements and objects were synchronized with the EEG headset, guaranteeing consistency and sustained tracking accuracy across experimental conditions.

### 3.2 Apparatus

Our study provides participants with an immersive experience using the Quest 2 virtual reality device, powered by a standard PC. We have also used HTC Vive Pro Eye in our ablation studies to show the performance in different VR hardwares. To ensure accurate and reliable data collection for EEG signals, participants were also connected to the DSI 24 EEG, a commercially available system that uses dry electrodes. This EEG system is fully compatible with our setup, ensuring good synchronization between the signals from the virtual reality headset.

### 3.3 Participants

We recruited 16 participants (6 female, 10 male) with ages ranging from 26 to 37 (M=25, SD=5.3) where the data were collected using an Institutional Review Board (IRB)-approved protocol. We obtained written informed consent from all participants while the study adhered to the standard of the Declaration of Helsinki. Each participant was asked to perform two experiments, after which they filled out a questionnaire, and when finished, they were given an $8 gift coupon. None of the participants had a history of severe motion sickness or cybersickness, indicating a low susceptibility. Moreover, individuals with photosensitive epilepsy were excluded from the study. During the experiment, participants were immersed in a virtual reality cycling simulation. The virtual bicycle's speed was set to simulate a pace of 15 miles per second, similar to real-world speed. At the beginning of the experiment, we also asked each participant whether the speed felt unrealistic (either too fast or too slow), but none expressed a desire to make any adjustments. The handlebar grip in the virtual environment was modeled at 40 cm in length, with the handlebar height adjusted to 100 cm to simulate a realistic cycling posture, ensuring participants felt as if they were on a real bicycle while maintaining ergonomic comfort. We also let participants move their heads freely in the VR environment during cycling, which allows the user to perform the task in various positions such as moving forward in the environment while looking forward or sideways.

### 3.4 Data Collection Procedure

The study comprised two sessions conducted on the same day. The experimenter briefed participants about the study's purpose and procedure in the first session. Then, the participants underwent calibration of the eye-tracker process by the manufacturer. The calibration involved following dots in the VR headset with the eyes for 30 seconds. Participants relaxed in VR for two minutes, first in an empty virtual environment (a minute) and then in the virtual environment of the experiment without any movements or interaction. The first session, lasting 35 minutes with two breaks of five to 10 minutes each, was followed by exposure to the second session for an additional 20 minutes. After the session, participants were asked to relax for at least 15 minutes during the study, and the experimenter reminded them not to operate machinery or drive a car within the subsequent hours.

#### 3.4.1 Joystick Controller

Different controllers were employed in VR environments [42, 44] to measure patients' cybersickness levels in their study. This unobtrusive method allowed participants to easily report the intensity of their symptoms during gameplay without being distracted, minimizing interference with the virtual experience. During our experiments, we followed a similar setup to [42] that allowed for continuous reporting of sickness levels. Specifically, the level of motion sickness was continuously reported by each subject using a joystick with a scale that ranged from 0 to 1, with discrete values of 0.1. The subjects were asked to raise or lower the scale whenever they felt more or less sickness, respectively. The scale was not shown to the subjects during the experiments to avoid any feedback loop. Three participants involved in the experiments did not indicate any degree of sickness on either joystick feedback and SSQ responses. At the end of the experiments, the sickness values were verified using the SSQ and correlated.

Using the joystick controller, we continuously obtain each participant's real-time sickness levels. An important advantage of this setup is that it shows variations in sickness levels among participants, even when they observe the same sequence of frames in VR. This observation presents a major drawback for previous works that attempt to detect cybersickness solely based on visual frames, as they overlook individual differences in how participants experience sickness, even when viewing the same content.

Second, obtaining real-time sickness levels allows us to observe sudden changes in physiological signals. For example, we observed that when participants changed the joystick level twice, it resulted in a notable increase in sickness level, whereas a single-level change had a comparatively lesser impact. Therefore, we encode participants' responses to ensure that their cybersickness severity remains consistent between consecutive segments if the change in response value is less than 10%, corresponding to a single-level shift.

This observation and threshold were further verified through the SSQ, where it was found that a single-level change did not induce any level of sickness among the participants. It is also important to note that everyone's perception of sickness level differs; in other words, the response values are relative to each individual. Therefore, this thresholding approach is to find the relative differences between responses while accounting for the inherent subjectivity in how participants experience and express cybersickness.

## 4 METHOD FOR CONTINUOUSLY ESTIMATING CYBERSICKNESS

### 4.1 Pre-processing

We use the EEG signals from the DSI-24 and inertial measurements from the VR device to detect cybersickness. Before, preprocessing the signals, we segmented them in 3-second windows without overlap. We employed a 3-second window to achieve a balance: longer windows might overlook sudden changes [51] in brain signals [4,45], and shorter ones would increase processing overhead without us observing a performance improvement with better temporal resolution. We also evaluated the performance with 2- and 4-second window lengths, but observed no significant improvements. We, then, process the EEG data to eliminate noise and artifacts. First, we resample it from 300 Hz to 100 Hz using an FIR antialiasing lowpass filter. We, then use a fourth-order bandpass Butterworth filter with a 1–40 Hz cut-off frequency for denoising. Lastly, we implement a notch filter at 50 Hz with a quality factor of five to eliminate power line interference from the recordings.

### 4.2 Modified Multitaper Spectrum Estimation

After preprocessing, we applied a modified multitaper method tailored for cybersickness detection to compute the power spectral density (PSD) of EEG signals in the range of 0.5 Hz to 40 Hz, which we outline in Section 4.2.1. Generally, the multitaper method outperforms periodogram techniques like Welch [65] for estimating the PSD of EEG signals as it diminishes temporal variability, ensuring a consistent spectrum estimation [63]. Additionally, the multitaper method averages modified periodograms obtained through mutually orthogonal tapers, which helps to reduc non-stationary noise in EEG signals that might occur when participants move in VR.

Before explaining our tailored multitaper PSD estimation (Section 4.2.1), we first provide a derivation of the standard multitaper PSD estimation for completeness, and notational consistency. Multitaper spectral density estimation provides better spectrum than the widely applied short-time Fourier transform methods thanks to the discrete prolate Slepian sequences [59] (DPSS). The DPSS arise from the following spectral concentration problem. Obtaining the discrete time Fourier transform (DTFT) $(X(f))$ of a finite time series $x[n]$, for which a sequence maximizes the ratio given in Equation 1, subject to the constraint that the sequence has finite power (Equation 2).

$$\lambda = \frac{\int_{-W}^{W} |X(f)|^2 df}{\int_{-F_s/2}^{F_s/2} |X(f)|^2 df} \quad (1) \qquad \int_{-F_s/2}^{F_s/2} |X(f)|^2 df < \infty, \quad (2)$$

where $F_s$ is the sampling rate of the signal $x[n]$ and $W$ is the interested frequency range. This ratio determines an index-limited sequence with the largest proportion of its energy in the frequency band $[-W,W]$ where it leads to the eigenvalue problem given in Equation 3.

$$\sum_{m=0}^{N-1} \frac{\sin(2\pi W(n-m))}{\pi(n-m)} g_k(m) = \lambda_k(N,W) g_k(n), \quad (3)$$

where $\lambda_k$ is the eigenvalues, and $g_k(n)$ is the DPSS values that correspond to $k$th Slepian sequence. The eigenvectors of this equation, $g_k(n)$, are the DPSS values, which are mutually orthogonal to each other. After obtaining DPSS values, the periodograms are calculated in

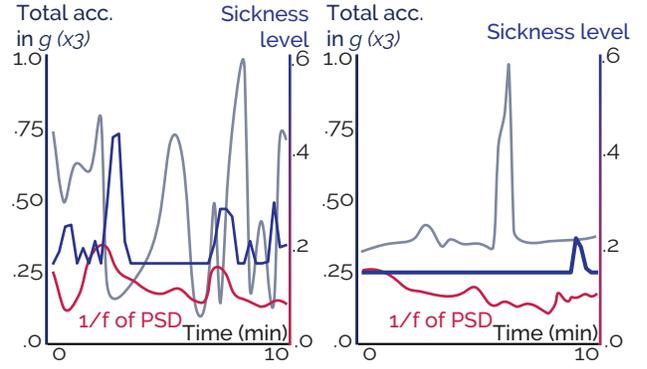

(a) An example of a user who experiences severe cybersickness during VR immersion

(b) An example of a user who does not experience cybersickness during VR immersion

Fig. 3: The plots illustrate the continuous variation in cybersickness levels (blue line; secondary y-axis) within the VR environment, showing a low correlation with head motion, as represented by the total acceleration (gray line; primary y-axis). The maximum value of sickness level changes from 0.1 to 0.85 between people. Figure 3b shows the temporal changes in sickness level for a participant during the experiment.

Equation 4 using a different Slepian sequence for each window, where we set the frequency resolution to 1 Hz.

$$S_k(f) = \Delta t \left| \sum_{n=0}^{N-1} g_k(n) x(n) e^{-j2\pi f n \Delta t} \right|^2 \quad (4)$$

Here $S_k(f)$ is the specific periodograms, each obtained using a different Slepian sequence $(g_k(n))$. Finally, the multitaper PSD estimate is calculated, by averaging all the periodograms using Equation 5.

$$S(f) = \frac{1}{K} \sum_{k=0}^{K-1} S_k(f) \quad (5)$$

#### 4.2.1 Temporal-relative PSD

The multitaper method is primarily designed for stationary random processes [63], which limits the duration of the analysis window for cybersickness. Given that cybersickness can change abruptly in a VR environment [31], applying the multitaper method to longer time frames becomes impractical due to the non-stationary nature of the process. To address this, we compute the PSD in 3-second windows and then calculate the difference between the current PSD and the average PSD of the initial three segments. Our method, which we term temporal-relative PSD, allows for detection of sudden changes in cybersickness while considering the temporal changes in EEG.

Our approach enables the model to learn the changes over time rather than focusing on absolute values, which hold little significance in the context of cybersickness while changing from subjects to subjects [18]. Moreover, since the multitaper calculation window is 3-seconds, it also enables us to calculate PSD without violating the stationary condition. In our ablation study, we also compare our proposed temporal-relative PSD estimation (TR-PSD) with the standard PSD estimation method outlined above.

#### 4.2.2 Spectral slope as a correlate of cybersickness

To demonstrate the continuous temporal changes in EEG signals during cybersickness in VR environments, we investigated the 1/f spectral slope [12,40] and, for the first time, established its link to cybersickness. The 1/f spectral slope, which we will simply refer to as **1/f**, describes the change of frequency power in the band of 30 Hz to 45 Hz assessed via the slope of a regression line fitted to the PSD on a log-log scale from 30–45 Hz (see [40] for further details). Studies have demonstrated that 1/f reflects the non-oscillatory, scale-free component of neural activity and effectively distinguishes wakefulness from diminished arousal levels [40]. We investigated the evolution of 1/f throughout participants' interaction in VR. As hypothesized, we observed that the 1/f, which is found to be related to arousal levels in humans, also links to cybersickness. For example, Figure 4 shows the extracted 1/f feature and the motion level with the reported continuous sickness severity of users during the VR task. Although the motion pattern is largely random, the users' sickness levels show a strong correlation (r=0.75±0.10) with the 1/f feature. Also, since the extracted 1/f feature stems from the power spectrum density, it is more robust towards motion artifacts than raw EEG signals as visible also in Figure 4. This suggests that EEG provides valuable insights into understanding cybersickness compared to inertial measurements–even though they remain commonly used.

### 4.3 Estimating Cybersickness

To estimate users' cybersickness levels in VR environments, we primarily used two estimators. After calculating the power spectral density using the proposed method, we input the TR-PSD along with kinematic data into a neural network model to detect cybersickness.

For the kinematic data, we extracted the following features without any filtering, similar to previous works [41]. We computed the first difference between two consecutive segments of the head position and the head rotation (i.e., x, y, z, roll, pitch, and yaw). We also computed the Euclidean norm of the head position and the head rotation to extract more information about the head movements in the VR environment. We then computed the first difference of the head speed and rotation. The overall algorithm with the processing steps are given in Algorithm 1. Lastly, we feed these extracted kinematic and EEG features together to the designed neural network.

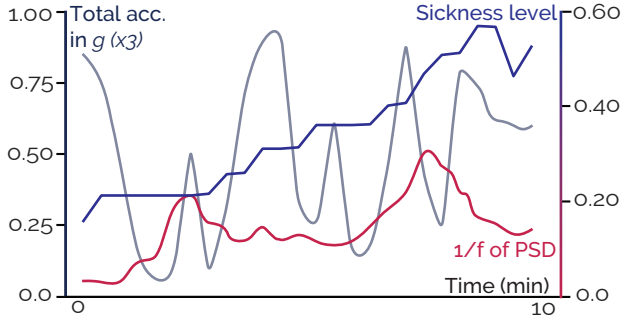

Fig. 4: An example showing total acceleration (gray line; primary y-axis), 1/f values extracted from processed PSDs (red line; secondary y-axis, right), and the user's cybersickness level (blue line; secondary y-axis) during a 10-minute VR immersion.

#### 4.3.1 Architecture

We design a neural network architecture that includes one encoder for each modality: one for EEG and one for inertial measurements. Each encoder is used to extract modality-specific features from the inertial measurements and spectral density of EEG signals independently before concatenating them to estimate the sickness level of participants. The input to our neural network architecture is processed following Algorithm 1 for both the EEG signals and the inertial measurements. Each encoder consists of a convolutional layer of 16 kernels of size of 2 and an LSTM cell with a hidden size of 32. We apply batch normalization [21] after each convolutional layer. We concatenate the features extracted from the two encoders and pass them through three fully connected linear layers with sizes of 10, 20, and 40, respectively. We use the ReLU activation function between linear layers, with a sigmoid activation applied at the end to predict the participant's real-time sickness response based on the PSD and kinematic features. We also give a schematic illustration of our architecture in Figure 5.

#### 4.3.2 Training

We used the Adam optimizer [37] with $\beta_1 = 0.9$, $\beta_2 = 0.999$, and a mini-batch size of 8. The learning rate is initialized to 0.0001 and

**Algorithm 1** Processing of signals with TR-PSD extraction

**Require:** EEG signals $x_n$ and inertial measurements $g_n$ of recording $i$, and discrete prolate sequences $\xi$.
**Ensure:** Chunked Power Spectral Density $\bar{\rho}_i$
1: $x'_i \leftarrow$ butterworth$(x_i)$
2: $\rho'_i \leftarrow \log_{10}\left(\left|\sum_{n=0}^{N-1} \xi_n x'_i e^{-j2\pi f n \Delta t}\right|^2\right)$ ▷ Multitaper PSD calculation
3: **for** idx **in** enum_windows(1, N, window_size=3s) **do**
4:     idx$_+$ ← next_window(idx)
5:     idx$_-$ ← prev_window(idx)
6:     $\lambda, \alpha \leftarrow g_i[\text{idx}]$ ▷ $\lambda$ is linear and $\alpha$ is angular motion 3-vectors
7:     $s \leftarrow \|\lambda\|_2^2$ ▷ linear speed i.e. $\sqrt{x^2+y^2+z^2}$
8:     $w \leftarrow \|\alpha\|_2^2$ ▷ angular speed i.e. $\sqrt{r^2+p^2+ya^2}$
9:     $\lambda_-, \alpha_- \leftarrow g_i[\text{idx}_-]$
10:     $s_- \leftarrow \|\lambda_-\|_2^2$
11:     $w_- \leftarrow \|\alpha_-\|_2^2$
12:     $k_i^{(\text{idx})} \leftarrow$ 16-tuple $\begin{bmatrix} \lambda & \alpha & s & w \\ (\lambda - \lambda_-) & (\alpha - \alpha_-) & (s - s_-) & (w - w_-) \end{bmatrix}$
13:     $\bar{\rho}_i^{(\text{idx})} \leftarrow \frac{1}{3}(\rho'_i[\text{idx}_+] + \rho'_i[\text{idx}] + \rho'_i[\text{idx}_-]) - \frac{1}{3}(\rho'_i[2] + \rho'_i[1] + \rho'_i[0])$
14: **end for**
15: $k_i \leftarrow \{k_i^{(\text{idx})}\}_{\text{idx}=0}^{\#indices}$
16: $\bar{\rho}_i \leftarrow \{\bar{\rho}_i^{(\text{idx})}\}_{\text{idx}=0}^{\#indices}$

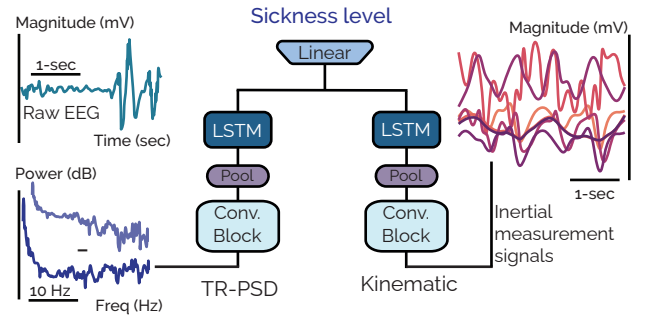

Fig. 5: The used ConvLSTM model architecture for predicting the cybersickness level in virtual reality. The proposed architecture has two inputs, the difference in PSD and the kinematic features, and one output for the sickness level.

reduced by half when the validation loss stops improving for 10 consecutive epochs. The training continues until 15 successive epochs without validation performance improvements. The best model is chosen as the lowest L1 loss on the validation data.

It is important to note that during our experiments, we strictly follow the leave-one-subject-out cross-validation approach to evaluate the model's ability to generalize. This ensures that if the models overfit the training data, their performance will decline on unseen testing subjects. This ensures robust evaluation and minimizes overfitting risks associated with the training data. This technique is advantageous in real-world scenarios because our model does not require person-specific data for calibration or training of the models. We trained the models three times for each left-out participant with different seeds and computed the performance as the average across them.

### 4.4 Model evaluation

We compare our method with existing models designed for cybersickness detection. We further investigate the impact of various components in our proposed method, with a particular focus on investigating how the choice of applying spectral density estimation and kinematics input influences performance. To evaluate the performance of these approaches, we have used multiple metrics to analyze their ability to detect and estimate participants' sickness levels within the VR environment. First, we used two commonly used regression metrics to investigate how well the models follow the sickness level continuously in real time. The continuous sickness holds significant importance as it provides insight into users' real-time cybersickness levels, distinguishing our approach from previous techniques that often require prolonged input before making a decision about the sickness level [35]. This real-time tracking capability is crucial for addressing and mitigating cybersickness, enhancing user experience, and the overall effectiveness of VR systems. These metrics are defined as follows:

$$\text{MSE} = \frac{1}{M} \sum_{i=1}^{M} (s_i - \tilde{s}_i)^2 \quad (6)$$

$$\text{MAE} = \frac{1}{M} \sum_{i=1}^{M} |s_i - \tilde{s}_i|, \quad (7)$$

where $M$ is the total number of segments of a participant in testing data, $s_i$ is the response on the controller in the $i^{th}$ segment about feeling symptoms of sickness, $\tilde{s}_i$ is the model prediction for the sickness level.

Furthermore, we calculated the accuracy of the models to evaluate if they can detect specific periods of cybersickness changes. Thus, we treated the participants' controller response as a binary category, where the patient is considered to be sick when it is above one degree in a neighborhood of a window, which is calculated as below:

$$\text{Acc} = \frac{TP + TN}{TP + TN + FP + FN}, \quad (8)$$

where TP represents the number of segments where both the user's joystick response and the model's prediction indicate sickness.

### 4.5 Real-time and SSQ Correlation Analysis

We computed the correlation between users' real-time response to the VR environment and their SSQs [55] that is collected at the end of the experiment to investigate the dependency between users' feelings of sickness and their embodiment in the virtual environment. Mainly, we investigated the Nausea, disorientation, and oculomotor-related subscores and the total score. In Table 1, micro-correlation is the correlation between each experiment conducted among thirteen users and macro-correlation is a user-wise correlation, where different experiments for each user are combined by taking the mean.

|  | micro-Correlation | macro-Correlation |
|---|---|---|
| Nausea | 56.9479% | 48.7619% |
| Oculomotor discomfort | 58.6206% | 56.3388% |
| Disorientation | 72.8282% | 75.5338% |
| Total | 69.8129% | 70.2778% |

Table 1: The correlation between embodiment and sickness

The significant correlation (p < 0.005) between the users' controller response and their overall reported feelings of sickness (nausea, oculomotor discomfort, and disorientation) holds valuable insights. It suggests that users who experienced more intense symptoms (higher scores on the questionnaire) also engaged with the joystick more frequently or intensely. This finding strengthens the validity of real-time cybersickness response measurement as a potential tool for capturing and quantifying the subjective experience of sickness during the experiments. Moreover, it establishes a direct link between verbal reports (questionnaire) and a behavioral measure (joystick activity), offering a more comprehensive understanding of users' experiences.

## 5 RESULTS AND DISCUSSION

We present the main results of our proposed approach compared to previous methods in Table 2. Overall, our proposed method demonstrates a substantial performance improvement, reaching up to 10–15%, in detecting and estimating the participants' cybersickness level. The results show that methods solely based on kinematic features fail to detect and estimate cybersickness levels. Furthermore, it is evident that the preprocessing step employed for extracting the power spectral density of EEG signals using the modified multitaper significantly improves performance. Notably, applying the spectral density estimation

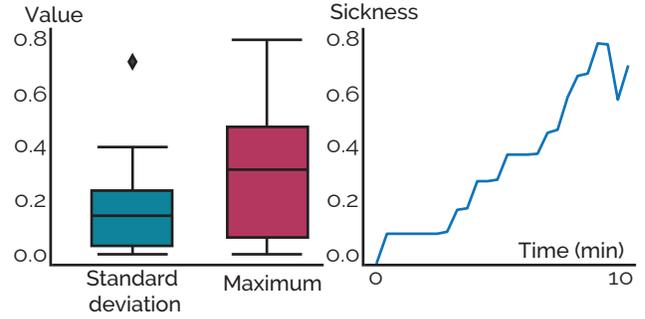

(a) The sickness distribution of users for the same frame of sequences

(b) An example of a continuous sickness change during VR immersion

Fig. 6: The plots show how the cybersickness level of participants changes continuously in the virtual reality environment. The maximum value of sickness level changes from 0.1 to 0.85 between people. Moreover, Figure 6b shows the continuous changes in sickness level for a participant during the experiment.

transformation increases the performance close to ≈ 20%, whereas incorporating information about the acceleration and rotation of the headset, commonly used in cybersickness detection literature, only increases approximately 5–6%. Moreover, adding the kinematics feature together with the raw EEG signals decreases the performance of the models. This implies that incorporating extra kinematic data might introduce noise or redundancy, particularly when the information comes from other modalities that do not explicitly relate to cybersickness, such as motion artifacts from raw EEG signals, leading to a decrease in the overall estimation performance.

### 5.1 Individual variation in cybersickness

We investigated the variation in cybersickness levels between participants in Figure 6. While all subjects were exposed to the same sequence of frames (i.e., videos) in the virtual reality environments, their reaction to the VR immersion and their cybersickness level differ significantly from person to person. For example, Figure 6a shows that the maximum sickness level reached by participants changes from 0.1 to 0.8 while the standard deviation of each session also reaches to 0.4. In other words, the participants' reaction to the VR immersion is dynamic and subjective. Therefore, although the sequence of frames can give information about the sensory conflict and the possibility of cybersickness level, it is not sufficient to detect subjective sickness differences across participants. This result also emphasizes that the same virtual environment can induce completely different responses in users for cybersickness, underscoring the importance of considering subjective measures, such as physiological signals, to accurately estimate the level of sickness continuously during VR immersion. Moreover, using the sequence of frames with a 3D convolutional neural network adds significant computational overhead to the overall monitoring system as the model needs to process a high dimensional input compared to power spectral densities or measurements from inertial units. This overhead can drain the battery lifetime of VR devices while decreasing the user experience significantly due to an increase in temperature [64].

### 5.2 Improvements from TR-PSD

During our experiments, we have also conducted several ablations to observe the contribution of each component in the performance of the overall method. One significant finding was that implementing our modified PSD calculation (temporal-relative PSD as in Algorithm 1) for sickness detection led to a performance increase of over 12% compared to the commonly-used straightforward employment of multitaper spectrum density estimation [61, 62]. This significant enhancement in performance underscores the pivotal contribution of our approach to cybersickness detection in VR.

Table 2: Performance comparison of our proposed method with ablations and prior works.

| Work | Input | Pre-Processing | Method | MAE ↓ | MSE ↓ | Acc (%) ↑ |
|---|---|---|---|---|---|---|
| Stereo-image | Frames | — | 3D ConvNets | 0.890 | 1.042 | 14.94 |
| Kinematic model [41] | IMU | Feature extraction from head IMUs | Conv-LSTM | 0.857 | 0.162 | 27.08 |
| Ours | IMU | IMU | Conv-LSTM | 0.931 | 0.193 | 38.22 |
| Ours | EEG | Filtering | Conv-LSTM | 0.841 | 0.182 | 44.32 |
| Ours | EEG | Filtering+PSD | Conv-LSTM | 0.751 | 0.143 | 58.97 |
| Ours | EEG | Filtering+TR-PSD | Conv-LSTM | **0.620** | 0.109 | 69.35 |
| Ours | EEG+IMU | Filtering+IMU | Conv-LSTM | 0.862 | 0.190 | 43.42 |
| Ours | EEG+IMU | Filtering+PSD+IMU | Conv-LSTM | 0.745 | 0.163 | 64.53 |
| Ours | EEG+IMU | Filtering+TR-PSD+IMU | Conv-LSTM | 0.638 | **0.092** | **76.83** |

\* TR-PSD refers to the proposed modified PSD calculation while considering the temporality in VR environments.

## 5.3 Window Length

We conducted an ablation study to investigate the impact of different window lengths for EEG signal processing on model performance. As shown in Table 3, the best performance was observed with a 3-second window, yielding a mean absolute error (MAE) of 0.638, a mean squared error (MSE) of 0.092, and an accuracy of 76.83%. Extending the window length to 5 seconds provided a slight improvement in MAE (0.632), but overall performance decreased, particularly in terms of MSE (0.143) and accuracy (75.25%). When using window lengths beyond 5 seconds, the model's performance further deteriorated, as indicated by a decrease in accuracy and higher error metrics at a 10-second window. These results suggest that shorter windows (around 3 seconds) are optimal for capturing the temporal dynamics of cybersickness in VR environments, whereas longer windows may fail to account for the rapid changes in users' symptoms.

Table 3: Ablation study on window length for EEG signals

| Window Length (seconds) | MAE | MSE | Acc (%) |
|---|---|---|---|
| 1 | 0.711 | 0.783 | 57.03 |
| 3 | 0.638 | **0.092** | **76.83** |
| 5 | **0.632** | 0.143 | 75.25 |
| 10 | 0.675 | 0.394 | 62.13 |

## 5.4 Different VR Hardware

To evaluate the system's performance across different VR hardware, we conducted an additional experiment with one subject using the HTC Vive Pro 2 headset. Following the same experimental procedure, we trained the same model from random initilization using data from the other subjects and tested it on the this subject. The cybersickness detection accuracy reached 72%, demonstrating the system's capability to generalize across different VR hardware devices as well. Our model relies on IMU signals from the VR device. However, for the EEG data, we used the same device in all experiments. Future work can investigate the performance of the system when using different EEG devices to further evaluate its generalizability.

## 6 SYSTEM OVERHEAD

Our algorithm reduces the dimensionality of the data and, subsequently, also the computational overhead by representing raw EEG signals as power spectral densities. Further, our approach reduces the number of modalities needed for cybersickness detection, enhancing both the efficiency and user experience of VR systems.

We evaluate the memory footprint and energy consumption of our method on the EFM32 Giant Gecko ARM Cortex-M3-based 32-bit microcontroller (MCU). For the multitaper implementation, we follow the approach described in [10], adjusting the taper length and parameters to suit our specific application. It has a 1024 kB flash and 128 kB of RAM with CPU speeds of up to 48 MHz, possessing minimal features.

Table 4 shows execution time, energy consumption, and required memory for each operation on our designed system. The operations are implemented and deployed to the target device using MATLAB (Coder Toolbox Release R2022b, The MathWorks, Inc, USA).

The overall execution time for an EEG segment, which is the segment length we use during our experiments takes 246 ms with 3.3 mJ energy consumption. Our proposed method requires a minimum RAM capacity of 128 KB, when the memory is allocated separately for each process. Since the overall execution time is less than the input segment, the proposed system enables continuous monitoring of the cybersickness level in real-time. Consequently, our method ensures high performance while adhering to the resource efficiency requirements, in terms of energy and memory.

## 7 BENEFITS, LIMITATIONS, AND FUTURE WORK

Our designed experiments with the proposed method shed light on multiple questions regarding the cybersickness in VR environments. First, our findings underscore the significance of considering both subjective and objective measurements when estimating and detecting cybersickness in virtual reality environments. Specifically, our results suggest that EEG signals are informative about the sickness level of users when the right pre-processing steps are applied to eliminate the noise and motion artifacts. In contrast, the stereo-image and only kinematics-based techniques can fail to detect and estimate sudden changes in sickness levels, especially considering the subjective responses of users to the VR. Thus, we believe that there is a huge room in future systems to design and prioritize different modalities, catering to diverse VR interaction needs. Additionally, advancements in VR technology should aim to address challenges associated with inside-out tracking, such as self-occlusion and limited tracking accuracy, to ensure consistent and reliable user experiences across different VR setups.

Furthermore, our study highlights the potential of neurophysiological markers as valuable indicators of cybersickness levels in VR environments. By correlating cybersickness with established neurophysiological markers, we bridge the gap between the effects of cybersickness on brain activity and existing research in neurophysiology. Our proposed method, leveraging a tailored EEG processing technique, demonstrates superior accuracy and reliability in continuously estimating cybersickness levels, paving the way for more effective interventions and improved user well-being in VR experiences.

Moreover, our comprehensive evaluation and ablation studies show the efficacy of our proposed method, surpassing current state-of-the-art approaches while decreasing the sensor modalities for a better battery lifetime and enhanced user experience. By openly sharing our analysis, results, and dataset, we contribute to the advancement of reproducible and generalizable research in the field of cybersickness detection.

Table 4: Memory footprint, execution time, and energy consumption of the system components.

| Operations | Exe. Time (ms) | Avg. Energy ($\mu J$) | Flash Memory Footprint (KB) | RAM Memory Footprint (KB) |
| --- | --- | --- | --- | --- |
| Filtering EEG | 18.2 | 2.03 | 2.41 | 9.67 |
| Multitaper PSD | 25.3 | 303 | 6.91 | 6.7 |
| Linear speed calc. | 3.01 | 92.3 | 4.43 | 32.4 |
| Angular speed calc. | 3.85 | 98.6 | 7.9 | 19.3 |
| PSD NN | 112 | 1.5k | 490 | 95 |
| IMU NN | 83.4 | 1.4k | 312 | 67 |
| Overall | 245.76 | 3395 | $\leq$ 512 KB | $\leq$ 128 KB |

Our study presents a significant step towards real-time cybersickness detection in VR using EEG and head motion. However, limitations offer valuable insights for future work. The relatively small sample size (N=16) with a narrow age range necessitates further research with larger and more diverse participant pools to investigate potential correlations between cybersickness and demographic factors, including gender, age, and potentially other characteristics [46]. Additionally, the controlled virtual environment limits the method's evaluation in more complex VR scenarios that might influence cybersickness differently. While our approach utilizes specialized EEG processing techniques, exploring advanced feature extraction and selection methods holds promise for enhanced cybersickness detection accuracy, potentially leading to the identification of true and general markers of susceptibility. Finally, the publicly available dataset offers a valuable resource for further exploration of cybersickness detection using EEG and head motion data, enabling researchers to refine existing methods or develop entirely new approaches, particularly those focused on understanding influences.

A major limitation of our study lies in the use of relatively limited virtual environments. To address this, future studies should aim to conduct experiments across a broader spectrum of virtual environments, encompassing various levels of complexity, interaction modalities, and sensory stimuli. By exploring cybersickness detection in diverse virtual settings, including object manipulation and diverse movement modalities, we believe that the importance of neurophysiological markers can be understood more comprehensively while refining the detection and monitoring methods accordingly for developing better intervention techniques. For example, in our study, we included a baseline measurement during the initial non-interaction phase of VR, where participants were passively exposed to the environment without active interaction. However, incorporating non-VR EEG measurements could provide a valuable comparison to better understand VR-induced cybersickness through neurophysiological markers. Additionally, investigating the impact of environmental factors such as visual fidelity, interactivity, and spatial layout on cybersickness susceptibility can provide valuable insights for designing more comfortable and immersive VR experiences. Overall, expanding the scope of experimental environments will contribute to the development of robust and adaptable cybersickness detection techniques that cater to the diverse needs and preferences of VR users.

An important aspect of our study is the 10–15-minute break between sessions, which included a relaxation period. This duration was chosen based on findings suggesting it is generally sufficient for recovery from cybersickness [68]. However, this duration may not fully eliminate residual fatigue or cybersickness in all participants. Future studies could explore the use of longer breaks or scheduled sessions on separate days to further minimize carry-over effects.

## 8 CONCLUSION

We have introduced a novel processing approach tailored to extracting meaningful features from EEG signals in the VR environment, which enables continuous estimation of cybersickness. Unlike traditional methods, our approach of processing EEG and head motion signals leverages a multitaper-based technique to achieve superior spectral resolution and reduced variance to eliminate noise and motion artifacts. We have shown that this results in more accurate and reliable predictions of cybersickness levels. We have also established a valuable connection between participants' dynamic cybersickness levels and physiological markers that are commonly used in neurophysiology, which shed light on the effect of cybersickness on brain activity. Through multiple evaluations, including ablation studies on our recorded cybersickness dataset, we have demonstrated that our method outperforms current state-of-the-art methods for cybersickness detection and estimation. Importantly, our method enhances the overall user experience and battery time by reducing the number of sensor modalities required and the computational demand.

To foster future research with a method focus in the challenging and important domain of cybersickness detection, we make our complete analysis, results, and models available publicly. We also release our dataset to support better opportunities for replication and validation of our findings by the broader research community. This way, we hope to accelerate advancements in cybersickness detection and support developments that may eventually even be capable of anticipating and preventing cybersickness and its symptoms from occurring in Virtual Reality and Mixed Reality more broadly, making these platforms available for prolonged use to a broader part of the population.

### ACKNOWLEDGMENTS

We thank all participants of our study. This work was supported in part by the European Union ERA-NET + EJP CHIST-ERA 2020 research and innovation program (GENESIS, CHIST-ERA-20-BCI-003 via SNSF 20CH21 203980).